\documentclass[%
reprint,
superscriptaddress,
 amsmath,amssymb,
 aps,
pre,
]{revtex4-2}

\usepackage{graphicx}
\usepackage{dcolumn}
\usepackage{bm}
\usepackage{hyperref,xcolor}
\hypersetup{colorlinks=true, allcolors=blue,urlcolor=blue}
\usepackage[mathlines]{lineno}

\usepackage{natbib}
\usepackage{tikz}
\usepackage{comment}
\usepackage{xcolor}

\def \fullin{\tikz[baseline=-0.6ex] {
\fill (0,0) circle (1.6pt) coordinate (i);
\node at (0,0)[left]{\small $i$};
\fill (2.2ex,0) circle (1.6pt) coordinate (j);
\fill (4.2ex,0) circle (1.6pt) coordinate (i');
\node at (4.2ex,0)[right]{\small $i'$};
\draw (i)--(j);
\draw(j)--(i');}
}

\def \copiedout{\tikz[baseline=-0.6ex] {
\fill (0,0) circle (1.6pt) coordinate (i);
\node at (0,0)[left]{\small $i$};
\fill (2.2ex,0) circle (1.6pt) coordinate (j);
\fill (4.2ex,0) circle (1.6pt) coordinate (i');
\node at (4.2ex,0)[right]{\small $i'$};
\draw (i)--(j);
\draw (j) (i');}
}

\def \origout{\tikz[baseline=-0.6ex] {
\fill (0,0) circle (1.6pt) coordinate (i);
\node at (0,0)[left]{\small $i$};
\fill (2.2ex,0) circle (1.6pt) coordinate (j);
\fill (4.2ex,0) circle (1.6pt) coordinate (i');
\node at (4.2ex,0)[right]{\small $i'$};
\draw (i) (j);
\draw (j)--(i');}}

\newcommand{\myname}{Dario Borrelli}
\newcommand{\myemail}{dario.borrelli@unina.it}
\newcommand{\myaffiliation}{Theoretical Physics Div., University of Naples Federico II, I-80125, Naples, Italy}
\newcommand{\titleofthispaper}{Largest connected component in duplication-divergence growing graphs \\ with symmetric coupled divergence}

\begin{document}

\title{\titleofthispaper}
\author{\myname}
 \email{\myemail}
\affiliation{\myaffiliation}
\date{\today}

\begin{abstract}
The largest connected component in duplication-divergence growing graphs with symmetric coupled divergence is studied. Finite-size scaling reveals a phase transition occurring at a divergence rate $\delta_c$.  The $\delta_c$ found is close to the locus of zero in Euler characteristic of finite-size graphs known to reflect the proximity of the largest connected component transition. A close correspondence with the vanishing of a scaling relation exponent for moments of the vertex degree distribution is shown, with such a scaling relation that generalizes a known form for duplication-divergence model graphs. The role of non-interacting vertices in shaping this transition with their presence or absence in duplication is also considered through a particular relation which would result in the two cases being comparable. The findings have relevancy for bond percolation in these growing graph models.
\end{abstract}

\maketitle

\textit{Introduction---}Sequentially growing network models are paradigmatic for the understanding of how the structure of complex networks emerges \cite{krapivsky2000,dorogovtsev2000,albert2000,krapivsky2001,jin2001}, and for studying what principles underlying growth and evolution lead to the emergence of their structural characteristics \cite{callaway2001,vazquez2003b,lambiotte2016}. Among these models, duplication-divergence models are based on the growth principle of duplication, according to which a randomly chosen vertex $i$ is duplicated into a copy vertex $i'$ with the same edges of $i$; divergence refers to probabilistic loss of duplicate edges \cite{ispolatov2005}. Some sophistications extend duplication-divergence models by adding edges other than those that are duplicated through, e.g., dimerization (an edge between $i$ and $i'$) \cite{vazquez2003,sole2008,bhat2016,sudbrack2018}, mutation (edges bet{\tiny {\tiny }}ween $i'$ and other vertices in the graph) \cite{kim2002,sole2002,pastor2003}, vertex deletion \cite{farid2006}. Most of these graphs aim at modeling structural characteristics of biological networks (e.g., pairwise protein interaction networks \cite{kim2002,sole2002,vazquez2003,pastor2003}, gene networks \cite{bhan2002,csengun2006}), the world-wide-web \cite{kleinberg1999,kumar2000}, scientific citation networks \cite{newman2018}, online social networks grown by vertex copying \cite{krapivsky2005,bhat2016,lambiotte2016}. Duplication-divergence is also among network growth processes that may admit a limiting power-law dependence of the vertex degree distribution \cite{raval2003some}, and the emergence of proportional preference in the growth process (e.g., see \cite{barabasi2016,newman2018,dorogovtsev2022}), as in Ref.~\cite{barabasi1999}.

When the divergence process considers loss of duplicate edges of vertex $i'$, it is referred to as complete asymmetric divergence \cite{borrelli2025}. Yet, the divergence process can also affect duplicate edges of $i$ and $i'$ with different probabilities (asymmetric divergence) or with equal probability (symmetric divergence) \cite{borrelli2025}. In the latter case, one can distinguish between two cases: (a) coupled divergence: given a duplicate edge pair $\{ (i,j), (i',j) \}$, only one edge of this pair can be lost while the other is retained; (b) uncoupled divergence, both edges of the pair can be lost \cite{ispolatov2005b,sudbrack2018}. In Ref.~\cite{borrelli2025}, the divergence asymmetry rate $\sigma$ allows generalizing the complete asymmetric divergence ($\sigma=1$), and the coupled symmetric divergence ($\sigma = 1/2$), as well as intermediate configurations between these two limit cases \cite{borrelli2025}. The model of Ref.~\cite{borrelli2025} also considers the possibility of presence (with parameter $d=0$) or absence ($d=1$) of non-interacting vertices among vertices for possible duplication, yet including them in the graph when the divergence process yields a non-interacting vertex.

For some of these growing network models prior studies showed structural phase transitions. These transitions concern: the model with complete asymmetric divergence with divergence probability $\delta=1$ and non-zero mutation probability showing an infinite-order percolation transition \cite{kim2002}, reminiscent of a Berezinskii-Kosterlitz-Thouless transition \cite{krapivsky2004}; the network growth by vertex copying showing distinction between sparse and dense networks \cite{lambiotte2016,bhat2016}; the complete asymmetric divergence model with inclusion of non-interacting vertices and with dimerization rates (suggesting changes in the network topology signaling the presence of a largest connected component transition \cite{de2022}); the symmetric coupled divergence sophisticated through both dimerization and mutation (suggesting a sharp change in the average connected component size \cite{sole2020} for the models in Ref.~\cite{pastor2003} and Ref.~\cite{vazquez2003}).

In Ref.~\cite{de2022}, for the complete asymmetric divergence model with non-interacting vertices, zeros of the Euler characteristic of finite-size graphs were considered indicative for the largest connected component transition, with loci $\delta_{c,\xi}$ of such zeros close to critical values $\delta_c$ of divergence probability that may be expected from a percolation transition. In this respect, an understanding of these structural changes in the minimal model of duplication-divergence with symmetric coupled divergence has never been deepened before.

Here, the duplication-divergence model with symmetric coupled divergence is studied by focusing on the largest connected component transition through both the Euler characteristic of finite-size graphs to infer $\delta_{c,\xi}$, and finite-size scaling to infer $\delta_c$ and the scaling behavior of the relative size of the largest connected component, also considering how non-interacting vertices may change such transition loci. The Letter is organized as follows: the duplication-divergence model with symmetric coupled divergence is introduced as a special case of the model in Ref.~\cite{borrelli2025}, with emphasis on model assumptions and the vertex degree distribution; results are then shown and discussed; finally, concluding remarks summarize the main findings.

\begin{figure}
\includegraphics{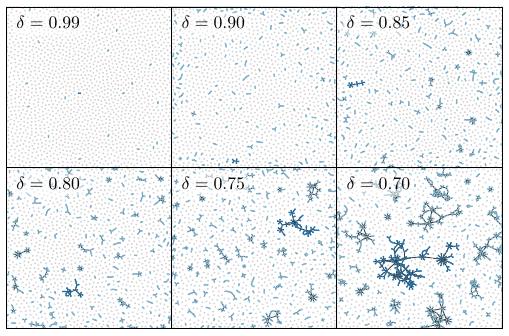}
\caption{Depiction of realizations of the generalized model in Ref.~\cite{borrelli2025} (with $\sigma=1/2$ and $d=1$) for various divergence rates $\delta$ (upper left of each panel). Each panel is a snapshot from a single simulation at $t = 2 \cdot 10^3$ vertices. The brightest dots are non-interacting vertices (vertex degree $k=0$), the darkest dots are vertices of the largest connected component. }
\label{fig1}
\end{figure}

\textit{Model---}For a coupled divergence process, in a growth iteration, each duplicate edge pair (i.e., $\fullin$) resulting from duplication has the following probabilities of transitioning to the configuration indicated in parentheses on the left-hand side of these equations
\begin{subequations}
\label{probs_coup}
\begin{gather}
 \mathcal{P}(\fullin) = \sigma(1-\delta) + (1-\sigma)(1-\delta) = 1-\delta,\label{probs_coup_a}\\
	 \mathcal{P}(\origout)= (1-\sigma)\delta,{\;\;} \mathcal{P}(\copiedout) = \sigma \delta,\label{probs_coup_b}
\end{gather}
\end{subequations}
with $\delta,\sigma \in [0,1]$ the divergence probability and the divergence asymmetry rate, from Ref.~\cite{borrelli2025}, with mutually exclusive events of edge loss from $i$ or $i'$. Eqs.~(\ref{probs_coup}) sum up to 1, and the corresponding events cover the entire sample space with intriguing symmetry properties.  When coupled divergence is symmetric, $\sigma = 1/2$ in (\ref{probs_coup_b}). 

A generic growth iteration $t$ can be enunciated as follows: a vertex $i$, chosen uniformly at random, either among interacting vertices (when $d=1$), or among all vertices including non-interacting ones (when $d=0$), is duplicated into a vertex $i'$ having the same edges of vertex $i$; for each pair $\lbrace (i,j), (i',j) \rbrace$ of duplicate edges, only one of the two edges of the pair is lost with probability $\delta$. This means that one chooses if to remove an edge of the pair (removal occurring with probability $\delta$), and where to remove it: either from $i$ with probability $1/2$ or from $i'$ with same probability $1/2$, but not from both. This divergence process is symmetric (because of an equal probability of losing an edge for $i$ and for $i'$), and also coupled [because removing the edge $(i,j)$ implies that the edge $(i',j)$ is not removed, and vice versa]. See Fig.~\ref{fig1}. 

For $d=1$, $k \geq 1$, asymptotically assuming linear scaling for the number of $k$-degree vertices with the number of interacting vertices (plausible for sparse graphs), the expected stationary vertex degree distribution $n_k$ reads
\begin{equation}
\hat{\mu} n_k = (1-\delta)[(k-1)n_{k-1} - kn_k] + \mathcal{M}_k,
\label{eq2}
\end{equation}
with $\mathcal{M}_k = 2\sum_{s=k}^{\infty} \binom{s}{k} q^{k} (1-q)^{s-k} n_s$, where $q$ may be considered equal to $(1-\delta)/2$, or $1-\delta/2$: the former emphasizes the concept of subfunctionalization in gene duplication that inspired the symmetric divergence scenario \cite{force1999preservation}; the latter reflects more formally the aforementioned growth iteration and it is the one considered hereafter (to deepen their relation, see Sec.~V of Supplemental Material \cite{supp}); resulting differences may be absorbed by a proper regularization.  Noticing that for large $k$ the summand of $\mathcal{M}_k$ is sharply peaked around $k/q$ and then $n_s$ is substituted with its value at $k/q$ \cite{krapvredn2003}; with these assumptions, (\ref{eq2}) admits a solution of the form $n_k \sim k^{-\gamma}$. To ensure consistency of Eq.~(\ref{eq2}) with the choice of $q=1-\delta/2$: $\hat{\mu} = 2\sum_{k}n_k[1-(\delta/2)^k]$. Having a normalized $n_k$ such that $\sum_{k\geq 1} n_k=1$, $\hat{\mu}$ can be written as $\mu + 1 = \sum_k n_k[1-2(\delta/2)^k]+1$, and (\ref{eq2}) becomes
\begin{equation}
(\mu+1) n_k = (1-\delta)[(k-1)n_{k-1} - kn_k] + \mathcal{M}_k,
\label{eq3}
\end{equation}
where $\mu$ is the expected rate of increase of interacting vertices (vertices with at least one edge). Given an initial graph with $t_0=2$ connected vertices, and being $t$ the total number of vertices (interacting plus non-interacting vertices) of the growing graph but also a discrete time variable counting the number of iterations ($\Delta t = 1$) as in Ref.~\cite{borrelli2025}, one has $\Delta N(\delta,t)/\Delta t=\sum_{k=1}^{\infty} n_k[1-(\delta\sigma)^k-(\delta - \delta\sigma)^k]$, being $(\delta\sigma)^k + (\delta - \delta\sigma)^k$ proportional to the probability of getting a non-interacting vertex either from $i'$ (first term of the sum) or from $i$ (second term of the sum); $\sigma=1/2$ yields $\Delta N(\delta,t)/\Delta t \propto \sum_k k^{-\gamma}[1-2(\delta/2)^k]$, which can give $\mu$ in (\ref{eq3}). When $t \gg t_0$, it turns out that the first term of the series is preponderant (see Supplemental Material \cite{supp}, Sec.~I) and, according to Ref.~\cite{ispolatov2005,borrelli2025}, the proportionality is substituted with a prefactor of $2$, yielding to consider in the large $k$ limit, the following 
\begin{equation}
N(\delta,t)\simeq 2(1-\delta)t,
\label{eq_tran}
\end{equation}
then $\mu := 2(1-\delta)$ is here assumed. With this $\mu$ in (\ref{eq3}), a transcendental equation relating $\gamma$ and $\delta$ reads
\begin{equation}
\gamma = 3 + \frac{1}{1-\delta} - \frac{2^{2-\gamma}}{1-\delta}(2-\delta)^{\gamma -1},
\label{eq5}
\end{equation}
which admits two branches of solutions, one is a trivial solution $\gamma=2$,  and a different non-trivial branch of solutions increasing monotonically with $\delta$, e.g., $\gamma=3$ at $\delta = 2- \sqrt{2}$, diverging asymptotically for $\delta$ approaching $1$. Note that, due to symmetry of sample space, regularization of (\ref{eq5}) yields $\gamma = 3 - 2^{2-\gamma}(1-\delta)^{\gamma-2}$ as in Ref.~\citep{borrelli2025} when $\sigma=1/2$, describing its essence without diverging terms that emerge when $\delta \rightarrow 1$.

\begin{figure}[b]
\includegraphics{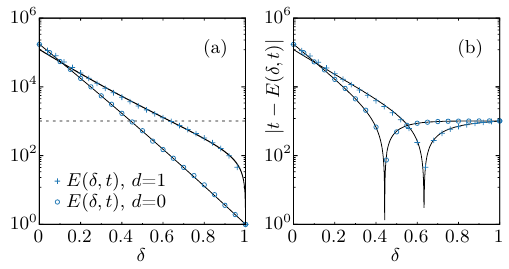}
\caption{In (a), plots of $E(\delta,t)$ for the model with $d=0$ and $d=1$ (see legend) with $t=1024$ (dashed horizontal line).  In (b), the respective $|t - E(\delta,t)|$ for the model with $d=0$ and for the model with $d=1$ showing a locus of singularity respectively at $\delta_{c,\xi} \approx 0.442$ and $\delta_{c,\xi} \approx 1 - e^{-1}$. Points shown are obtained through averaging over $10^3$ simulations; solid curves are obtained from (\ref{eq_et}) for $d=0$ and from the moment distribution for $d=1$.}
\label{fig2}
\end{figure}

Without stationarity assumption in the rate equation for $n_k$, one can begin by considering the expected number of vertices with degree $k$ when the graph has a total number of vertices $t$, denoted by $N_k(t)$, and the proportion of $k$-degree vertices as $n_k(t) = N_k(t)/t$, and then, $\frac{\partial N_k(t)}{ \partial t} = t\frac{\partial n_k(t)}{\partial t} + n_k(t)$. Being $N:=N(\delta,t)$ the number of vertices with degree $k\geq 1$, when $N \neq t$ but generically $N=\mu t$, it follows that $\mu \frac{\partial N_k}{\partial N} = N\frac{\partial n_k}{\partial N} + \mu n_k$, where here $n_k(N):=N_k(N)/N$ (with $k \geq 1$) defines the expected vertex degree distribution normalized such that $\sum_{k\geq 1} N_k= N$, and $\sum_{k\geq1} n_k=1$, due to variable change $t \rightarrow N$. A rate equation for the evolution of $n_k$ reads
\begin{equation}
\begin{gathered}
N\frac{\partial n_k}{\partial N} + \mu n_k = (1-\delta) \left[ (k-1)n_{k-1} -kn_k \right] + \mathcal{M}_k - n_k,
\end{gathered}
\label{rteqn}
\end{equation}
where $\mathcal{M}_k$ is that of (\ref{eq2}) with $q=1-\delta/2$. On the right-hand side of (\ref{rteqn}), when multiplied by $(1-\delta)$, the two terms in squared parentheses are respectively: (i) the contribution of adjacent vertices of a vertex with degree $k-1$ chosen for duplication when the duplicate edge is not lost (a gain term for $n_k$), (ii) the contribution of adjacent vertices of a $k$-degree vertex chosen for duplication when the duplicate edge is not lost (a loss term for $n_{k}$). $\mathcal{M}_k$  is a gain term for $n_k$ due to all possible removal of $s-k$ edges due to divergence, where $q^{k}(1-q)^{s-k}$ reflects the binomial choice with $k$ edges retained and $s-k$ edges lost; the factor $2$ in $\mathcal{M}_k$ accounts for such a removal process for $i$ and $i'$. The term $-n_k$ represents the situation in which vertex $i$, when initially has vertex degree $k$, disappears from the count of $k$-degree vertices, replaced by its current vertex degree. Note that $i$ and $i'$ are indistinguishable after duplication due to the coupled symmetric nature of the divergence process. One can also recast (\ref{rteqn}) via a continuum approximation
\begin{equation}
N\frac{\partial n_k}{\partial N} + (\mu +1) n_k = (\delta-1)\frac{\partial (kn_k)}{\partial k} + \mathcal{M}_k, 
\label{rteqn_rec}
\end{equation} 
which yields (\ref{eq3}) in the stationary limit assumption $\partial n_k / \partial N = 0$. 
With a non-stationary assumption, $l$-moments of $n_k$ are computed by summing over all $k$ and multiplying by $k^l$ (see Supplemental Material \cite{supp}, Sec.~IV), which yields 
\begin{subequations}
\begin{gather}
M_l(N,\delta) \propto N^{\Phi(l,\delta,\mu)}, \\
\Phi(l,\delta,\mu) = l(1-\delta) + 2(1-\delta/2)^l - (\mu + 1).
\end{gather}
\label{expmnt}
\end{subequations}
Eqs.~(\ref{expmnt}) are a generalized form of that provided in Ref.~\cite{vazquez2003,vazquez2003b} (which emerges exactly with $\mu=1$ where dimerization is also considered). The nonlinearity of $\Phi(l,\delta)$ reflects the multifractal nature \cite{vazquez2003b,dorogovtsev2002multifractal} of this model for the symmetric coupled duplication-divergence case when considering non-stationarity of $n_k$. The $\mu:=2(1-\delta)$ would yield a constant average vertex degree, and $M_2$ would diverge for $\delta<2-\sqrt{2}$, agreeing with the transcendental equation (\ref{eq5}) with $\gamma = 3$ at $\delta=2-\sqrt{2}$ found with the hypothesis of stationarity in the thermodynamic limit. A peculiar feature of $\mu$ is that it acts as a gauge fixing offset ensuring stationarity of $M_1$ (constant average vertex degree), with a relative shift of the order of moments (which are relative to $M_1$). Calculating $\partial \Phi / \partial l$, expanding the resulting nonlinear terms, and calculating the curvature $\frac{\partial}{\partial \delta ^2} (\partial \Phi / \partial l)|_{\delta \rightarrow 0} = l - 1/2$, suggests that fractional moments may be relevant, which is expected given the multifractality (indeed, one will see that $E(\delta,t)$ with $t$ fixed for $d=1$, shown in Fig.~\ref{fig2} for $t=1024$, is calculated through the $1.5$th moment).  

The mean number of edges $E(\delta,t)$ for the model with $d=0$, with an initial graph with two connected vertices ($t_0=2$, $E_0=1$) as in Ref.~\cite{borrelli2025} (see also Supplemental Material \cite{supp}, Sec.~II) reads as
\begin{equation}
E(\delta,t)  = E_0 \frac{\Gamma(t_0)  \Gamma[2(1-\delta)+t]}{\Gamma[t_0+2(1-\delta)] \Gamma(t)}, 
\label{eq_et}
\end{equation}
with $\Gamma(\cdot)$ the Euler Gamma function. Note that, for increasingly large $t$, the model with $d=0$ shows the growth pattern of $E(\delta,t)$ versus $N(\delta,t)$ of the model with $d=1$ [see inset of Fig.~S1 in Supplemental Material \cite{supp} (Sec.~I)], yet described by Eq.~(\ref{eq_et}) with $t$ (as shown in Ref.~\cite{borrelli2025}). The relevant difference between the model with $d=0$ and with $d=1$ stands in the mean number of edges given a fixed $t$, due to the non-uniform probability distribution of choosing a vertex for duplication, which is non-zero only for vertices with degree $k\geq 1$ when $d=1$.
\begin{figure}[t]
\includegraphics{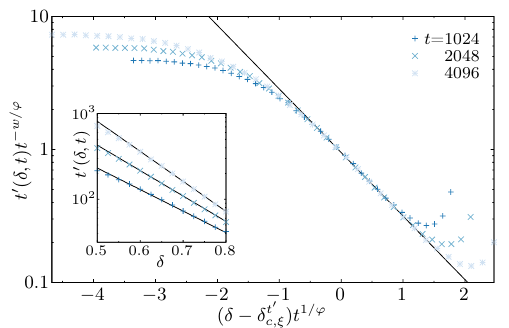}
\caption{Finite-size scaling for $t'(\delta,t)$ with scaling collapse for $\delta$ approximately in $[0.55,0.75]$. The linear behavior in log-linear plot suggests the exponential form (shown for visual reference), $a\cdot\mathrm{exp}[-b(\delta -\delta_{c,\xi}^{t'})t^{1/\varphi}]$ with $a \approx 0.95$, $b \approx 1.05$ which, when unscaled, describes a subset of points of $t'(\delta,t)$ vs. $\delta$ (see inset); points are averages over $3 \cdot 10^3$ simulations.}
\label{fig3}
\end{figure}
\begin{figure}
\includegraphics{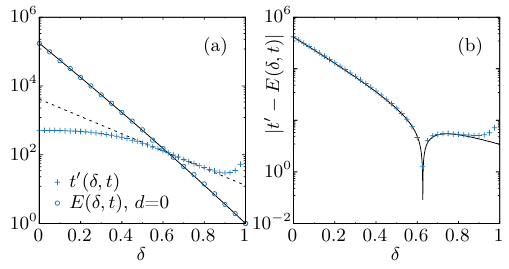}
\caption{In (a), $E(\delta,t)$ as in Fig.~\ref{fig2}(a), $t'(\delta,t)$ as in the inset of Fig.~\ref{fig3} with $t=1024$. In (b), Euler entropy curve assuming $t'(\delta,1024)$ in the model with $d=0$, with $\delta \in [0.55,0.75]$ from Fig.~\ref{fig3} extended for visual reference to $\delta \in [0,1]$; the locus of zero of $|t'-E(\delta,t)|$ is near the one for $d=1$ in Fig.~\ref{fig2}(d).}
\label{fig4}
\end{figure}

\textit{Euler characteristic---}In Ref.~\cite{de2022}, it has been shown that topological transitions can be found through the Euler characteristic of a simplicial complex, and for the special case of a duplication-divergence graph with complete asymmetric divergence ($\sigma=1$). The Euler characteristic is defined as $-\sum_{n}(-1)^{n}\kappa_{n}$, where $\kappa_{n}$ is the total number of $n$-cliques. Then, as in Ref.~\cite{de2022}, for the special case of a duplication-divergence graph, which does not include $n$-cliques with $n \geq 3$, the Euler characteristic assumes the form $t - E(\delta,t) $ (with $t$ the total number of vertices in the graph), and the Euler entropy is the natural logarithm of its absolute value, $\mathrm{ln}(|t - E(\delta,t)|)$. Thus, zeros of Euler characteristic correspond to singularities in Euler entropy, with a singularity locus $\delta_{c,\xi}$ arising at the formation of $n$-cycles and expected to be close to a critical value $\delta_c$, where the largest connected component transition occurs \cite{de2022}. For finite-sized graphs, Fig.~\ref{fig2} shows a $\delta_{c,\xi} \approx 0.442$ (see Supplemental Material \cite{supp}, Sec.~II) that agrees with the result in Ref.~\cite{de2022} for complete asymmetric divergence ($\sigma=1$) although here, instead, it is found for the symmetric coupled divergence case ($\sigma=1/2$) of the model with $d=0$. Euler entropy of finite-size graphs with $t=1024$ and $d=1$ is close to $\delta_{c,\xi} \approx 1-e^{-1}$ [see, Fig.~\ref{fig2}(b)]. Due to a different number of non-interacting vertices, (\ref{eq_tran}) may not directly hold for the model with $d=0$, nonetheless, the mean number of edges of interacting vertices can be matched across the case of $d=0$ and $d=1$ (see inset of Fig.~S1 in Supplemental Material \cite{supp}), and one can get the corresponding $t$ (as if the model was with $d=1$) from the model with $d=0$, through the following relation (with $\delta \neq 1$)
\begin{equation}
t'(\delta,t) \simeq N(\delta,t)  / 2(1-\delta).
\label{eq_trnsf}
\end{equation}
Note that $t'$ is a $\delta$-dependent function and not a fixed value as in the calculation of the Euler entropy shown in Fig.~\ref{fig2}. It is suggested that $t'(\delta,t)$ has also a scaling form for increasing $t$ in a subset of $\delta$ values where curves for various $t$ collapse on the same function. The ansatz considered here for such a scaling of $t'(\delta,t)$ is
\begin{equation}
t'(\delta,t)=t^{w/\varphi} h \left[ (\delta-\delta_{c,\xi}^{t'})t^{1/\varphi}\right],
\end{equation}
with $w$, $\varphi$, $h(\cdot)$ respectively two unknown exponents and a scaling function. Different curves for various $t$ collapse on the same curve for $\delta \in [0.55,0.75]$ approximately (see Fig.~\ref{fig3}), with $\delta_{c,\xi}^{t'} = 0.638 \pm 0.006$, $w = 2.831 \pm 0.095$, $\varphi = 4.177 \pm 0.367$, which is indeed the range of $\delta$-values where $\delta_{c,\xi}$ is expected in the model with $d=1$. Intriguingly, the transformation $t'(\delta,t)$ on the model with $d=0$ can be leveraged to get an Euler entropy curve different from the one estimated numerically for the model with $d=1$, yet exhibiting nearly the same singularity locus of the Euler entropy for finite-sized graphs, e.g., $\delta_{c,\xi}^{t'} \approx \delta_{c,\xi}$ with $d=1$, $t=1024$, see Fig.~\ref{fig4}(b), Fig.~\ref{fig2}(b). The value of $\delta_{c,\xi}$ considers finite-sized graphs and it may be only indicative of the largest connected component transition, thus, in the following, finite-size scaling for the relative size of the largest connected component is tackled to consider $\delta_c$ in infinite-sized graphs from finite-size behavior \cite{newman1999monte,radicchi2010explosive}.

\begin{figure}[t]
\includegraphics{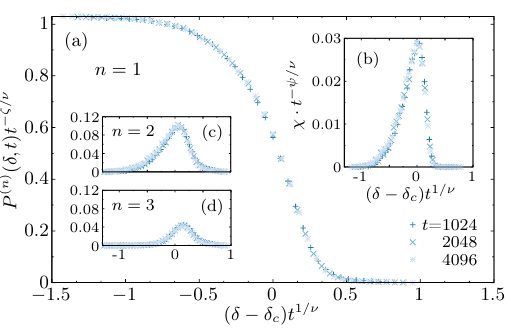}
\caption{Scaling collapse of $P_\infty(\delta,t):= P^{(n=1)}(\delta,t)$ in (a), and $\chi(\delta,t)$ in (b), respectively from Eq.~(\ref{eq_sclP}) and Eq.~(\ref{eq_sclChi}). Scaling collapse for $P^{(n=2)}(\delta,t)$  and $P^{(n=3)}(\delta,t)$ respectively shown in (c) and (d), see also Sec.~III in Supplemental Material \cite{supp}; all points shown were obtained by averaging over $3 \cdot 10^3$ simulations.}
\label{fig5}
\end{figure}

\textit{Finite-size scaling---}An ansatz for the probability of a vertex to belong to the $n$-th largest connected component $P^{(n)}(\delta,t)$, being $P_{\infty}:=P^{(n=1)}$ for the largest (with $n=2$ the 2-nd largest, $n=3$ the 3-rd largest,...) is written
\begin{equation}
P^{(n)}(\delta,t) = t^{\zeta/\nu} f^{(n)}\left[ (\delta-\delta_c)t^{1/\nu} \right],
\label{eq_sclP}
\end{equation}
where $\zeta$, $\nu$, $f^{(n)}(\cdot)$ are respectively two unknown exponents and a scaling function on which one would expect scaling collapse of $P^{(n)}(\delta,t) $ for various sizes $t$, i.e., various graph order in the language of graph theory. Denoting $P_{\infty}:=P_{\infty}(\delta,t)=P^{(1)}(\delta,t)$, and its susceptibility
\begin{equation}
\chi(\delta,t) = t \left( \langle P_{\infty}^2 \rangle - \langle P_{\infty} \rangle^2 \right),
\end{equation}
which characterizes the intensity of fluctuation about the mean of the order parameter, then the following scaling ansatz for the susceptibility is written
\begin{equation}
\chi(\delta,t) = t^{\psi/\nu} g\left[ (\delta-\delta_c)t^{1/\nu}\right],
\label{eq_sclChi}
\end{equation}
with $\psi$ an unknown exponent, and $g(\cdot)$ a scaling function.
For the largest connected component, the value $\delta_c$ here found is close to the locus $\delta_{c,\xi}$: in Fig.~\ref{fig5}, the collapse on the same curve is obtained for $\zeta = -0.033 \pm 0.006$, $\nu = 9.634 \pm 0.069$, and $\delta_c= 0.6  \pm 0.002$ (formal uncertainties affected by discretization effects of $\delta$ with $\Delta \delta =0.025$), and $\psi$ determined by satisfying the  relation $\psi/\nu = 1 + 2\zeta / \nu$ (scaling is in terms of ``volume", total number of vertices $t$). Note that, from Eqs.~(\ref{probs_coup}), $p=1-\delta$ would link what here studied to a bond percolation on growing graphs while the exponents may be reminiscent of those of an explosive transition with trivial exponents $\zeta=0$ and $\psi/\nu =1$ according to relations between exponents in \cite{radicchi2009explosive,radicchi2010explosive}, with some plausible analogy to jamming \cite{piscitelli2021jamming}; yet, the estimated exponent $\zeta$ is non-zero and of the order of $10^{-2}$ (extremely small), thus only $\zeta \approx 0$ and $\psi/\nu \approx 1$, which also suggests considering this transition as continuous \cite{lee2011expl,da2010explosive}.

\begin{table}[t]
\caption{Values of $\delta$ (exact for $l=2$, numerical otherwise) at which $\Phi(l,\delta)=0$ ($l$ is in the column header), with $\mu = 2(1-\delta)$ in Eqs.~(\ref{expmnt}); when $\mu=1$ in (\ref{expmnt}) one has a $\Phi=0$ exactly at $\delta=1/2$ and $\delta=4-2\sqrt{3}$ respectively for $l=1$ and $l=2$.}
\label{tab1}
\begin{ruledtabular}
\renewcommand{\arraystretch}{1.25}
\begin{tabular}{lcccccc}
 & $l=1.5$ & $l=2$ & $l=2.5$ & $l=3$ & $l=3.5$ \\
\colrule
$\delta$ & 0.562 & $2-\sqrt{2}$ & 0.610 & 0.635 & 0.661 \\
\end{tabular}
\end{ruledtabular}
\end{table}

\begin{figure}[!b]
\includegraphics{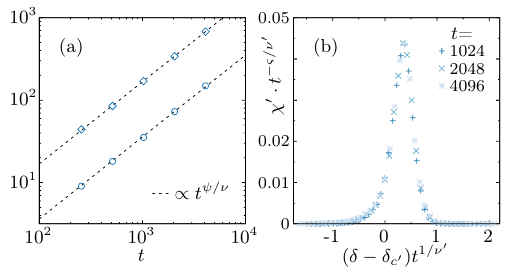}
\caption{In (a): scaling with $t$ of $\chi^{*}$ ($\diamond$) and $\langle s \rangle^{*}$ ($\circ$) for increasing sizes $t$: $256,512,1024,2048,4096$. Dashed lines are visual references for the scaling $t^{\psi/\nu}$. In (b): scaling collapse of $\chi'(\delta,t)$, with $\delta_{c'}$ that precedes $\delta_{c,\xi}$ shown in Fig.~\ref{fig2}(b) for $d=0$; $\chi'$ quantifies the fluctuation about $\langle s' \rangle$.}
\label{fig6ab}
\end{figure}

The critical value $\delta_c$ is found within the region where exponents of $n_k$ moments vanish (see Table~\ref{tab1}). As shown in Fig.~\ref{fig6ab}(a), $\psi/\nu$ describes the scaling of peaks $\chi^{*}$ of $\chi(\delta,t)$, for various $t$, and $\langle s \rangle^{*}$ of the weighted average connected component size $\langle s \rangle$, the latter defined as $\langle s \rangle \propto \sum_{s=1}^{\infty} s^{2}C_s(\delta,t) - s_{\infty}^2$, with $C_s(\delta,t)$ and $s_{\infty}$ respectively the expected number of connected components of size $s$ and size of the largest connected component. Indeed, at the critical point, one would expect $\chi(\delta_c,t) \sim t^{\psi/\nu}$ (see also Supplemental Material \cite{supp}, Sec.~III), provided that the estimates of $\psi$ and $\nu$ are correct \cite{newman1999monte}. As discussed for the Euler entropy, the role of non-interacting vertices appears to be crucial in changing the locus of the critical value in topological transitions of the duplication-divergence graph model with symmetric coupled divergence. From such a consideration, one can further define the following observable $\langle s' \rangle = \sum_{s>1} s C_s(\delta,t)/\sum_{s>1}C_s(\delta,t)$,  being it an unconventional average connected component size in which non-interacting vertices ($s=1$) have not been considered in the sum. Then, one can consider (with $\delta \neq 1)$ $\chi'(\delta,t) = (\langle s'^2 \rangle -\langle s' \rangle^2)/ N(\delta,t)$, proportional to the fluctuation about the mean quantity $\langle s' \rangle$. Then for $\chi'$, a different scaling ansatz is written
\begin{equation}
\chi'(\delta,t)= t^{\varsigma/\nu'} \ell\left[ (\delta -\delta_{c'})t^{1/\nu'}\right],
\end{equation}
with $\varsigma$, $\nu'$ two unknown exponents and $\ell(\cdot)$ a scaling function on which one may expects collapse of curves for various $t$ with the correct choice of the exponents and of $\delta_{c'}$. The estimated exponent is $\nu' = 6.185 \pm 0.002$, with the scaling collapse occurring for $\varsigma= 6.827 \pm 0.002$, see Fig.~\ref{fig6ab}(b). The value of $\delta_{c'} = 0.425 \pm 0.001$, slightly anticipates the value $\delta_{c,\xi} \approx 0.442$ found for the model with $d=0$. Noteworthy is that the estimated $\delta_{c'}$ precedes $1/2$ which it may occur presumably with, e.g., slow dimerization, mutation rates \cite{sole2020}, yet it is also worth noting that non-interacting vertices have a relevant role in shaping the transition studied. While in the model with $d=1$ at the critical point $\delta_c$ there might be a scaling $\langle s\rangle \propto t^{\psi/\nu}$ [Fig.~\ref{fig6ab}(a)], further research would further characterize it as the focus here was mainly on the largest connected component. Cases with general divergence asymmetry rates $\sigma$ are also yet to be deepened.

\textit{Conclusion---}The largest connected component in the duplication-divergence network model with symmetric coupled divergence ($\sigma=1/2$) plausibly emerges at a divergence rate $\delta_c$ found via finite-size scaling with close correspondence with a here generalized non-stationary (multifractal) form of vertex degree distribution moments, nearly agreeing with loci of zeros of Euler characteristic in finite-size graphs. Non-interacting vertices presence or absence in duplication was suggestive to change loci, making them comparable via a proper relation. The findings contribute to enhancing knowledge of evolving networks with multiple connected components. 

\begingroup
\makeatletter
\let\oldaddconentsline\addcontentsline
\renewcommand{\addcontentsline}[3]{}
\makeatother

\bibliography{apssamp}

\clearpage
\newpage

\makeatletter
\let\addcontentsline\oldaddcontentsline
\makeatother
\endgroup

\onecolumngrid  

\setcounter{equation}{0}
\setcounter{figure}{0}
\setcounter{table}{0}
\setcounter{page}{1}
\setcounter{section}{0}

\makeatletter
\renewcommand{\theequation}{S\arabic{equation}}
\renewcommand{\thefigure}{S\arabic{figure}}
\renewcommand{\thetable}{S\arabic{table}}
\renewcommand{\thepage}{p\arabic{page}}
\renewcommand{\large}{\@setfontsize\large{12pt}{14pt}}
\makeatother

\begin{center}
\large{\textbf{Supplemental Material for:\\``\titleofthispaper"}}\\ \vspace{0.15in}

\normalsize{\myname$^{1, *}$ }\\ \vspace{0.05in}

{\small $^1$\textit{\myaffiliation}} \\ \small{(Dated: \today)}
\vspace{0.15in}
\end{center}

{\insert\footins{\footnotesize\textsuperscript{*}\href{mailto:\myemail}{\myemail.}}}

\def\tocname{CONTENTS} 
\tableofcontents 

\vspace{0.1in}

\section{Continuum approach, $N(\delta,t)$}
\label{apndx}

Here it is shown that (4) (in the main text) can be approached through a continuum approximation, comparing it with simulations. Let $C_s(\delta,t)$ be the expected number of connected components of size $s$ when the growing graph with divergence rate $\delta$ has a total number $t$ of vertices.Let $N(\delta,t)$ indicates the expected number of interacting vertices. This quantity is equivalent to the number of $k$-degree vertices with $k\geq 1$, which can be written as
\begin{equation}
N(\delta,t)=\int_{s>1}C_s(\delta,t)ds.
\end{equation}
The total number of vertices $t$ is instead
\begin{equation}
t = \int_{s>1}C_s(\delta,t)ds + C_1(\delta,t) = \int_{s\geq 1}C_s(\delta,t)ds,
\end{equation}
where $C_1(\delta,t)$ is the number of non-interacting vertices, i.e., vertices with no edges. Then, the expected number of interacting vertices $N(\delta,t)$ also results from the total number of vertices $t$ by subtracting  $C_1(\delta,t)$, i.e.
\begin{equation}
N(\delta,t) = t - C_1(\delta,t).
\label{3}
\end{equation}
Now, let $N_k(\delta,t)$ be the expected number of vertices with degree $k$ when the growing graph -- with divergence rate $\delta$ and divergence asymmetry rate $\sigma$ -- has a total number $t$ of vertices, and let one denotes with $n_k$ the expected fraction of vertices with degree $k$. Then, the rate at which the number of interacting vertices increases with $t$ can be written as
\begin{equation}
\frac{\partial N(\delta,t)}{\partial t} = \int_{k\geq1} n_k\left[ 1-(\sigma\delta)^{k}  - (\delta-\sigma\delta)^k \right] dk.
\end{equation}
\noindent If one assumes an expected fraction of $k$-degree vertices $n_k \sim k^{-\gamma}$, the rate at which $N(\delta,t)$ increases with $t$ can be written as
\begin{equation}
\frac{\partial N(\delta,t)}{\partial t} = \mathcal{C}\int_{k\geq1} k^{-\gamma}\left[ 1-(\sigma\delta)^{k}  - (\delta-\sigma\delta)^k \right]dk,
\label{rt_n}
\end{equation}
with $\mathcal{C}$ a proportionality factor; (\ref{rt_n}) can be conveniently rewritten as 
\begin{equation}
\frac{\partial N(\delta,t)}{\partial t} =\mathcal{C}(1-\delta) + \mathcal{C}\int_{k>1} k^{-\gamma}\left[ 1-2(\delta/2)^{k}  \right]dk.
\end{equation}
In the large $k$ limit, with a prefactor 2 (from \cite{ispolatov2005,borrelli2025})
\begin{equation}
\frac{\partial N(\delta,t)}{\partial t} \simeq 2(1-\delta),
\end{equation}
which yields (with $\delta \neq 1$)
\begin{equation}
N(\delta,t) \simeq 2(1-\delta)t.
\label{A8}
\end{equation}
When instead one considers $\delta \rightarrow 0$, thus  $C_1(\delta \rightarrow 0,t) \rightarrow 0$ due to a slow divergence rate which reduces the probability of generating a non-interacting vertex through duplication-divergence, then
\begin{equation}
N(\delta \rightarrow 0,t)=t,
\label{A9}
\end{equation}
which follows directly from (\ref{3}). This theoretical result is compared with numerical simulations in Fig.~\ref{figA1} for finite-sized graphs; for increasing size $t$, simulations show a behavior that approaches the theoretical prediction of (\ref{A8}) and (\ref{A9}). What shown here indeed was a continuum approximation that holds for increasing large values of the total number of vertices $t$ (that includes both interacting vertices and non-interacting vertices).

\begin{figure}[!b]
\includegraphics{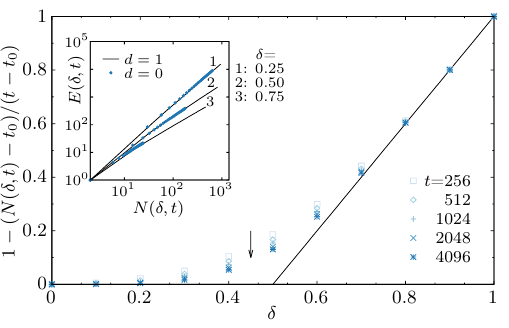}
\caption{Expected proportion of non-interacting vertices versus $\delta$ for various $t$ (legend of main plot). Each point is an average over $3 \cdot 10^3$ simulations. The arrow highlights that, with increasing $t$, points from simulations would approach the theoretical behavior: solid line is obtained through $N(\delta,t)$ of Eq.~(\ref{A8}), whose slope was suggested in Ref.~\cite{borrelli2025}; when $\delta \rightarrow 0$, simulations follow the behavior that results from Eq.~(\ref{A9}). Inset: growth of $E(\delta,t)$ versus $N(\delta,t)$ for $d=0,1$ (see legend) and various $\delta$ (see numbered legend). Points are averages over $10^2$ simulations ended at $t=10^3$.}
\label{figA1}
\label{figA1}
\end{figure}

\section{$E(\delta,t)$, and Euler Characteristic}
\label{apndxB}
As in Ref.~\cite{de2022}, Euler entropy of graphs considered here is the logarithm of the absolute value of the Euler characteristic $t-E(\delta,t)$. Eq.~(9) (in the main text) provides an analytic form for the expected number of edges $E(\delta,t)$. Singularities in Euler entropy correspond to zeros of the Euler characteristic, occurring for $t = E(\delta,t)$. The recurrence equation for $E(\delta,t)$ can be rewritten as
\begin{equation}
E(\delta,t+1) = E(\delta,t) \left( \frac{2-2\delta+t}{t} \right).
\end{equation}
Starting from $t_0=2$, one can begin to explicit the first few iterations
\begin{equation}
\begin{gathered}
E(\delta,t_0=2)=1, \hspace{0.1in} E(\delta,3) = \left( \frac{2-2\delta+2}{2} \right), \hspace{0.1in} E(\delta,4) = E(\delta,3) \left(\frac{2-2\delta+3}{3}\right)=\left(\frac{2-2\delta+2}{2}\right)\cdot\left(\frac{2-2\delta+3}{3}\right). \\
\end{gathered}
\label{Et_patt}
\end{equation}
Following the same pattern, one can continue writing Eqs.~(\ref{Et_patt}) until a generic iteration $t$, yielding
\begin{equation}
\begin{aligned}
E(\delta,t) = \left( \frac{2-2\delta+2}{2} \right) \cdot \left( \frac{2-2\delta+3}{3} \right) \cdot \left( \frac{2-2\delta+4}{4} \right) \dots \left( \frac{2-2\delta+t-1}{t-1} \right),
\end{aligned}
\end{equation}
which can be recast as
\begin{equation}
E(\delta,t) = \frac{(2-2\delta+t-1)\cdot(2-2\delta+t-2)\dots(2-2\delta+2)}{(t-1)\cdot(t-2)\dots 2 \cdot 1}.
\end{equation}
By using factorial notation, it follows
\begin{equation}
E(\delta,t) = \frac{(1-2\delta+t)!}{(t-1)!(3-2\delta)!}.
\label{exact_withfactorial}
\end{equation}
The factorial form of the Euler's Gamma function $\Gamma(x)=(x-1)!$ is leveraged to recast (\ref{exact_withfactorial}), yielding
\begin{equation}
E(\delta,t) = \frac{\Gamma(2 -2\delta +t)}{\Gamma{(t)}\Gamma{(4-2\delta)}}.
\label{Et_mfex}
\end{equation}
Then, one can consider the series expansion at $t \rightarrow \infty$ of Eq.~(\ref{Et_mfex}), which gives
\begin{equation}
E(\delta,t) = \frac{t^{2(1-\delta)}}{\Gamma(4-2\delta)} + \frac{t^{1-2\delta}(2\delta^2 - 3\delta +1)}{\Gamma(4-2\delta)} + \dots
\label{serexp}
\end{equation}
Considering (\ref{serexp}) with the first two terms of the series, for $t-E(\delta,t)=0$ one finds a $\delta_{c,\xi} \approx 0.4421094\dots$, with $t=1024$.

\section{Scaling of the second and the third largest connected component}
\label{apndxC}

Analogously to the scaling argument for the largest connected component, one can consider the scaling ansatz for the second largest connected component, the third largest connected component.  Denoting the mean relative size of the second largest connected component as $P^{(2)} := N^{(2)}(\delta,t)/t $, i.e., the ratio between the mean number of vertices in the second largest connected component $N^{(2)}(\delta,t)$ and the total number of vertices $t$ of a growing graph by duplication-divergence (symmetric coupled\textcolor{blue}{,} $\sigma=1/2$) with divergence probability $\delta$.  Similarly, for the third largest connected component, $P^{(3)} := N^{(3)}(\delta,t)/t$. With the same critical exponents $\zeta,\nu$ found for $P_{\infty}$, recalling the scaling ansatz
\begin{equation}
P^{(n)}(\delta,t) = t^{\zeta/\nu}f^{(n)}\left[ (\delta-\delta_c)t^{1/\nu} \right],
\label{sa_n2n3}
\end{equation}
\noindent where here $n=2,3$, and $f^{(1)}(\cdot),f^{(2)}(\cdot)$ are respectively the scaling functions for the relative size of second and third largest connected component on which one expects scaling collapse of different curves $P_{\infty}^{(n)}(\delta,t)$ for various size $t$, provided the proper choice of exponents $\nu,\zeta$ and the critical value $\delta_c$.  Recalling the exponents found for $P^{(1)}(\delta,t)$ (the largest connected component): $\nu=9.634 \pm 0.069$, $\zeta = -0.033 \pm 0.006$, $\delta_c = 0.6 \pm 0.002$. With this choice of $\nu,\zeta,\delta_c$, insets (c) and (d) of Fig. 5 (in the main text) show scaling collapse when plotting $P^{(n)}(\delta,t)$ versus $(\delta-\delta_c)t^{1/\nu}$; with $n=2$ (second largest connected component), scaling collapse is on the scaling function $f^{(2)}(\cdot)$ in Fig.~5(c) (main text). With $n=3$ (third largest connected component), scaling collapse is on the scaling function $f^{(3)}(\cdot)$ in Fig.~(5)(d) (main text). Remarking that $p=1-\delta$ would map to a bond percolation on growing graphs for which standard notation of exponents would be $\beta:=\zeta$ in Eq.~(14) (main text), $\gamma:=\psi$ in Eq.~(16) (main text), with proper adjustments.

\begin{figure}[t]
\includegraphics{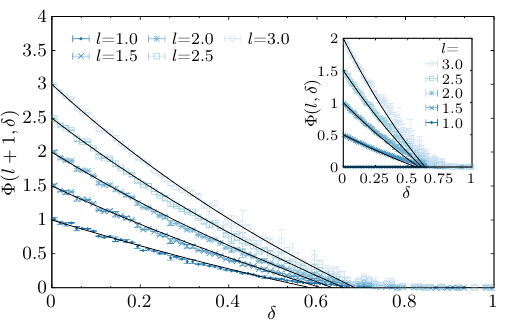}
\caption{The exponent $\Phi$ of the moments of $n_k$ for various moment indices $l$ (see legend); Points result from simulations, solid lines from Eq.~(\ref{Ml}). The inset shows $\Phi$ with moments relative to $M_1$ [stationary with $\mu=2(1-\delta)$] showing a shift by 1 in indices of moments. Moments were computed with size $t$ ranging from $1024$ to $4096$ vertices; $\Phi$ (of points shown) were estimated from fitting a powerlaw of the plot $\langle k^l \rangle$ vs $t$. }
\label{figA2}
\end{figure}

\section{Moments equation and scaling of $n_k$}
\label{appD}
Being $n_k$ the vertex degree distribution, and $M_l = \sum_k k^l n_k$ its $l$th-moment equation, the rate equation approach \cite{krapivsky2001,krapvredn2003} for the evolution of $n_k$ is recalled
\begin{equation}
\begin{gathered}
N\frac{\partial n_k}{\partial N} + n_k(\mu +1) = (1-\delta) \left[(k-1)n_{k-1} -kn_k \right] + \mathcal{M}_k,
\end{gathered}
\label{Meq1}
\end{equation}
where $\mathcal{M}_k = 2 \sum_{s \geq k} \binom{s}{k}q^k (1-q)^{s-k}n_s$, with $q=1-\delta/2$ and $\mu=2(1-\delta)$. Then, (\ref{Meq1}) is rearranged and multiplied both sides by the sum of over all $k$ and by $k^l$
\begin{equation}
\begin{gathered}
\sum\nolimits_k k^l \left[ N\frac{\partial n_k}{\partial N} \right] = \sum\nolimits_k k^l \biggl[(1-\delta) \left[(k-1)n_{k-1} -kn_k \right] - n_k(\mu +1)+ \mathcal{M}_k \biggr].
\end{gathered}
\label{Meq2}
\end{equation}
The left hand side is $N\partial M_l / \partial N$. On the right hand side of (\ref{Meq2}), one term is
\begin{equation}
\begin{gathered}
\sum\nolimits_k k^l \biggl[(1-\delta) (k-1)n_{k-1} -kn_k \biggr] =  (1-\delta)\sum_{r=1}^l \binom{l}{r-1} M_r \simeq l(1-\delta)M_l,
\end{gathered}
\label{Meq3}
\end{equation}
where a shift of indices of the sum has been used for the first term in squared parentheses in (\ref{Meq3}), and $(k+1)^l \simeq k^l + lk^{l-1}$ for the last $\simeq$ (namely, the term $r=l$). Then, having $\mu=2(1-\delta)$, the following term is trivial
\begin{equation}
\sum\nolimits_k k^l  \left[ n_k(\mu + 1) \right] = (\mu + 1) \sum\nolimits_k k^l  n_k = (\mu + 1)M_l.
\label{Meq4}
\end{equation}
The last term on right hand side of (\ref{Meq2}) is then
\begin{equation}
\sum\nolimits_k k^l \mathcal{M}_k =2 q^l M_l + l(l-1)q^{l-1}(1-q)M_{l-1} + \mathcal{O}(M_{l-2}). 
\label{Meq5}
\end{equation}
Putting these results together yields
\begin{equation}
\begin{gathered}
N\frac{\partial M_l}{\partial N} = \biggl[l(1-\delta) + 2q^{l} - \mu - 1 \biggr]M_l + \biggl[ l(l-1)q^{l-1}(1-q) \biggr]M_{l-1} + \mathcal{O}(M_{l-2}).
\end{gathered}
\end{equation}
When  $N \gg 1$, the following scaling for the $l$th-moment is considered
\begin{equation}
M_l \propto N^{l(1-\delta) + 2q^{l} - \mu - 1 },
\end{equation}
i.e.,  Eqs.~(8) in the main text; substituting $\mu=2(1-\delta)$ yields
\begin{subequations}
\begin{gather}
M_l \propto N^{\Phi(l,\delta)},\\
\Phi(l,\delta) = l(1-\delta) + 2q^{l} - 3 + 2\delta,
\end{gather}
\label{Ml}
\end{subequations}
\noindent with the nonlinear dependence on $l$ that suggests the multifractal nature of the distribution $n_k$  \cite{dorogovtsev2002multifractal,vazquez2003b}, in the non-stationary case. With $q=1-\delta/2$ and $\mu=2(1-\delta)$, reminiscent of a gauge fixing offset, one gets $M_1 \propto \mathrm{const.}$, due to the identity $(1-\delta) + 2q - 3 + 2\delta=0$. FIG.~\ref{figA2} shows $\Phi$ in a comparison with simulations. 

\section{$q=(1-\delta)/2$ and $q=1-\delta/2$}
\label{appE}
Consider the generalized model with arbitrary $\sigma$ and $\delta$ both defined in $[0,1]$. The sample space of these probabilities for the coupled divergence case is such that
\begin{equation}
1 = (1-\delta)(1-\sigma) + (1-\delta)\sigma + \sigma \delta + \delta (1-\sigma).
\label{eq9}
\end{equation}
Then, let denote the binomial sum weighting $n_k$ in $\mathcal{M}_k$ as $\mathcal{B}_k(q)$
\begin{equation}
\mathcal{M}_k(q) = \sum_{s \geq k} \binom{s}{k}q^{k}(1-q)^{s-k}n_s = \mathcal{B}_k(q)n_s.
\end{equation}
Note that, in the rate equation for $n_k$, the structure of $\mathcal{M}_k$ appears two times accounting for the $i$ and $i'$ divergence process
\begin{equation}
\mathcal{S}_k(q_i,q_{i'}) = \mathcal{M}_k^{(\sigma)}(q_i) +  \mathcal{M}_k^{(1-\sigma)}(q_{i'}) = \biggl[ \mathcal{B}(q_i) + \mathcal{B}(q_{i'})\biggr]n_s,
\end{equation}
where the second identity leverages the linearity of the binomial transformation. The probabilities $q_i$ and $q_{i'}$ are related to $\sigma$ and $\delta$, depending on the choice of success probability. Let one consider two cases $(a)$ and $(b)$
\begin{subequations}
\begin{align}
&q^{(a)}_i = \sigma(1-\delta), \hspace{0.2in} q^{(a)}_{i'} = (1-\sigma)(1-\delta), \label{12a}\\
&q^{(b)}_i = 1-\sigma\delta, \hspace{0.36in} q^{(b)}_{i'} = 1-(1-\sigma)\delta. \label{12b}
\end{align}
\end{subequations}
Making explicit (\ref{12b}) as a function of (\ref{12a}) and of the other probabilities in the sample space according to (\ref{eq9}), it yields
\begin{subequations}
\begin{align}
&q^{(b)}_i = q^{(a)}_i + q^{(a)}_{i'} + (1-\sigma)\delta, \\
&q^{(b)}_{i'} = q^{(a)}_i + q^{(a)}_{i'} + \sigma \delta,  \\
&q^{(b)}_i + q^{(b)}_{i'} =2-\delta, \\
&q^{(a)}_i + q^{(a)}_{i'} =1-\delta.
\end{align}
\end{subequations}
Hence, it follows that
\begin{equation}
\left[q^{(b)}_i + q^{(b)}_{i'}\right] - \left[q^{(a)}_i + q^{(a)}_{i'}\right] = 1.
\end{equation}
When considering the following sums $\mathcal{S}_k(q^{(a)}_i, q^{(a)}_{i'})$ and $\mathcal{S}_k(q^{(b)}_i, q^{(b)}_{i'})$, a symmetry with respect to the mean emerges
\begin{equation}
\langle \mathcal{S}_k(q^{(b)}_i, q^{(b)}_{i'}) \rangle = \langle \mathcal{S}_k(q^{(a)}_i, q^{(a)}_{i'}) \rangle + \langle n_s \rangle,
\end{equation}
being $\langle \mathcal{S}_k(q_i,q_{i'}) \rangle = (q_i +q_{i'}) \langle n_s \rangle$ due to linearity of the expected value. In terms of total mass, $\mathcal{S}_k(q^{(b)}_i, q^{(b)}_{i'})$ is a translated version of $\mathcal{S}_k(q^{(a)}_i, q^{(a)}_{i'})$ of a quantity exactly equal to the expected value of $n_k$. Hence, one can consider the rate equation for $n_k$ in a mean-field description using $q=(1-\delta)/2$ (in coupled symmetric divergence $\sigma=1/2)$ or more formally follow the generic growth iteration with $q=1-\delta/2$ and including the $n_k$, which yields in the stationary assumption the plus one term in the $(\mu + 1)n_k$ term of the left hand side of the balance equation of (\ref{Meq1}), yielding Eq.~(3) in the main text.


  
\end{document}